\documentclass[aps,prl,twocolumn,superscriptaddress]{revtex4-1}
\usepackage{graphicx}
\usepackage{amsmath}
\usepackage[dvipsnames]{xcolor}
\usepackage[normalem]{ulem}
\usepackage[colorlinks,linkcolor=blue,citecolor=blue,urlcolor=blue]{hyperref}
\usepackage{algorithm}
\usepackage{algorithmic}

\newcommand{\kr}[1]{{\bf q}\cdot{\bf r}_#1}
\newcommand{\jpm}{J_{\pm}}
\newcommand{\jz}{J_z}

\begin{document}


\title{Dynamics of Topological Excitations in a Model Quantum Spin Ice}

\author{Chun-Jiong Huang}
\author{Youjin Deng}
\email[]{yjdeng@ustc.edu.cn}
\affiliation{Shanghai Branch, National Laboratory for Physical Sciences at 
Microscale and Department of Modern Physics, University of Science and 
Technology of China, Shanghai, 201315, China}
\affiliation{CAS Center for Excellence and Synergetic Innovation Center in 
Quantum Information and Quantum Physics, University of Science and Technology 
of China, Hefei, Anhui 230026, China}
\affiliation{CAS-Alibaba Quantum Computing Laboratory, Shanghai, 201315, China}
\author{Yuan Wan}
\email[]{yuan.wan@perimeterinstitute.ca}
\affiliation{Perimeter Institute for Theoretical Physics, Waterloo, Ontario N2L 2Y5, Canada}
\affiliation{Institute of Physics, Chinese Academy of Sciences, Beijing 100190, China}
\author{Zi Yang Meng}
\email[]{zymeng@iphy.ac.cn}
\affiliation{Institute of Physics, Chinese Academy of Sciences, Beijing 100190, China}
\affiliation{CAS Center of Excellence in Topological Quantum Computation and School of Physical Sciences, University of Chinese Academy of Sciences, Beijing 100190, China}

\date{\today}

\begin{abstract}
We study the quantum spin dynamics of a frustrated \textit{XXZ} model on a 
pyrochlore lattice by using large-scale quantum Monte Carlo simulation and 
stochastic analytic continuation. In the low-temperature quantum spin ice 
regime, we observe signatures of coherent photon and spinon excitations in the 
dynamic spin structure factor. As the temperature rises to the classical spin 
ice regime, the photon disappears from the dynamic spin structure factor, 
whereas the dynamics of the spinon remain coherent in a broad temperature 
window. Our results provide experimentally relevant, quantitative information 
for the ongoing pursuit of quantum spin ice materials.
\end{abstract}


\maketitle

{\it Introduction} --- A prominent feature of quantum spin liquids (QSLs) is 
their ability of supporting topological excitations, i.e., elementary 
excitations whose physical properties are fundamentally different from those of 
the constituent spins~\cite{Balents2010,Savary2017}. Detecting topological 
excitations in dynamic probes, such as inelastic neutron scattering, nuclear 
magnetic resonance, resonant inelastic x-ray scattering, and Raman scattering 
probes, provides an unambiguous experimental identification for 
QSLs~\cite{Coldea2010,Han2012,Mourigal2013,Kimura2013,Banerjee2017,Sibille2017,Fu2015,ZLFeng2017,Pan2016,Tokiwa2016,
Ament2011,Lummen2008,Maczka2008,YuanWei2017}. Understanding the dynamics of 
topological excitations is therefore essential for interpreting experiments on 
QSL. While the dynamics of one dimensional QSL are well understood, thanks to a 
wide variety of available analytical and numerical tools~\cite{Giamarchi2003}, 
much less is known in higher dimensions. On the one hand, mean field 
approximations, although offering a crucial qualitative understanding of the 
topological excitations, are often uncontrolled for realistic spin 
models~\cite{Fradkin2013,Auerbach1994,Wen2007}. On the other hand, exactly 
solvable spin models are few and far 
between~\cite{Kitaev2006,Knolle2014,Changlani2017}. Therefore, an unbiased 
numerical approach such as quantum Monte Carlo (QMC) calculations stands out as 
a method of choice, as it can provide unique insight into the dynamics of QSLs 
in higher dimensions.

\begin{figure}[htp!]
\includegraphics[width=\columnwidth]{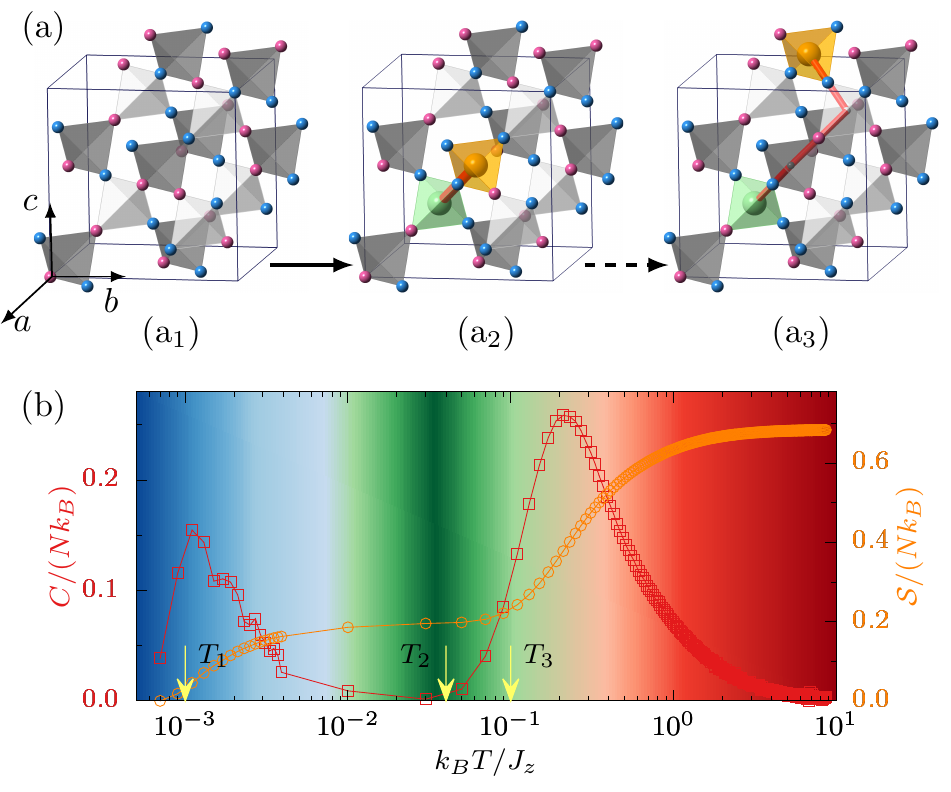}
\caption{(a) \textit{XXZ} model on pyrochlore lattice. Gray box shows a cubic 
unit cell along with the orientation of the cubic axes $a,b,c$. Small red and 
blue spheres denote $S^z=1/2$ and $-1/2$ states respectively. Starting from a 
spin configuration in the ice manifold (a$_1$), one may flip a spin and create 
a pair of spinons with charges $Q=1$ (gold sphere) and $Q=-1$ (light green 
sphere) residing on neighboring tetrahedra (a$_2$). The spinons may propagate 
in the lattice by flipping a string of spins (red solid line) (a$_3$). (b) 
Thermal entropy $\mathcal{S}$ (orange open circles, right vertical axis) and 
specific heat $C$ (red open squares, left vertical axis) as a function of 
temperature $T$. The regions corresponding to the trivial paramagnetic regime, 
the classical spin ice regime, and the quantum spin ice regime are shaded in 
red, green, and blue, respectively. Bright yellow arrows mark the temperatures 
at which we carry out QMC study.}
\label{fig:fig1}
\end{figure}

\begin{figure*}[htp!]
\includegraphics[width=\linewidth]{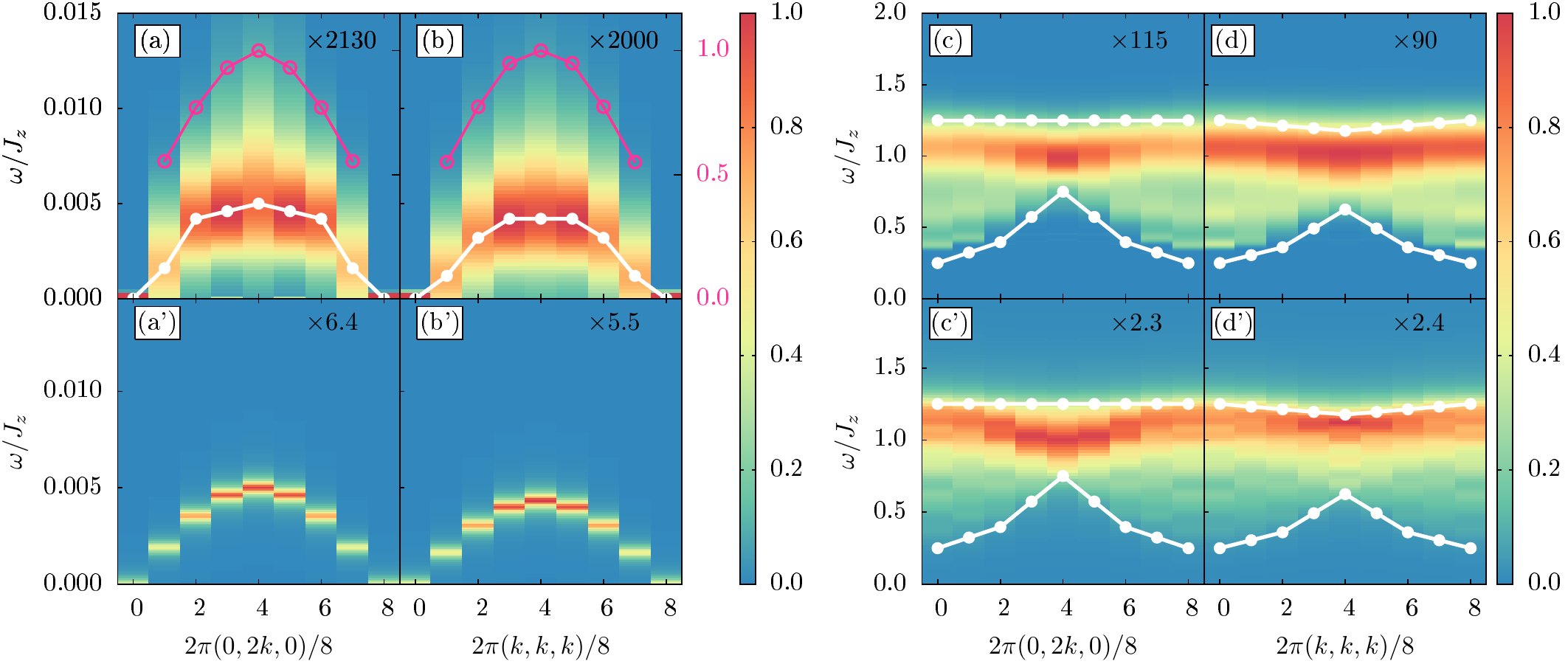}
\caption{Left panel: (a,b) Dynamic spin structure factor 
$S^{zz}(\mathbf{q},\omega) \equiv 
\sum_{\alpha}S^{zz}_{\alpha\alpha}(\mathbf{q},\omega)$ obtained from QMC-SAC at 
temperature $T_1$ along high symmetry cubic directions $(010)$ and $(111)$. The 
photon appears as a gapless branch of excitation with dispersion starting from 
Brillouin Zone center. White dots mark the position of spectral peaks. Pink 
open circles show the integrated spectral weight at each momentum point with 
maximal spectral weight rescaled to 1. (a',b') Photon spectra calculated from a 
Gaussian QED model. Right panel: (c,d) Dynamic spin structure factor 
$S^{+-}(\mathbf{q},\omega) \equiv 
\sum_{\alpha}S^{+-}_{\alpha\alpha}(\mathbf{q},\omega)$ obtained from QMC-SAC at 
$T_1$. The spectra show a dispersive continuum of two-spinon excitations. 
(c',d') The results from a tight-binding model calculation, where the spinons 
are modeled as free particles. The calculated spectra are then broadened with a 
Lorentzian to mimic interaction effects. The spinon continuum boundaries 
calculated from the tight-binding model are marked as white dots in both 
QMC-SAC spectra (c,d) and the theoretical spectra (c',d').}
\label{fig:fig2}
\end{figure*}

In this work, we study the dynamics of quantum spin ice (QSI), a paradigmatic 
example of three dimensional 
QSL~\cite{Bramwell2001,Hermele2004,Molavian2007,Ross2011,Onoda2011,Savary2012,Gingras2014}.
 In QSI, $S=1/2$ spins form a pyrcohlore lattice, a network of corner-sharing 
network of tetrahedra [Fig.~\ref{fig:fig1} (a$_1$)]. The dominant Ising 
exchange interaction in the global spin $\hat{z}$ axis energetically favors a 
large family of spin configurations collectively known as the ice manifold, 
where every tetrahedron of the pyrochlore lattice obeys the ice rule: $Q_\alpha 
\equiv \eta_\alpha \sum_{i\in\alpha} S^z_i = 0$ [Fig.~\ref{fig:fig1} (a$_1$)]. 
Here, $S^z_i$ is the $\hat{z}$ component of the spin on lattice site $i$, and 
the summation is over a tetrahedron $\alpha$. $\eta_\alpha = -1(1)$ if $\alpha$ 
is an up (down) tetrahedron. The other subdominant exchange 
interactions~\cite{Hermele2004} induce quantum tunneling in the ice manifold, 
resulting in a liquidlike ground state that preserves all symmetries of the 
system. Viewing $S^z_i$ as the electric field and the ice rule as Gauss's law 
in electrostatics~\cite{Henley2005}, the spin liquid ground state is analogous 
to the vacuum state of the quantum electrodynamics (QED)~\cite{Hermele2004}.

Three types of topological excitations can emerge from the QSI ground 
state~\cite{Hermele2004}: The \emph{photon}, analogous to the electromagnetic 
wave, is a gapless, wavelike disturbance within the ice manifold. The 
\emph{spinon} is a gapped point defect that violates the ice rule within a 
tetrahedron [Fig.~\ref{fig:fig1} (a$_2$)]. In the QED language, spinons are 
sources of the electric field; the charge carried by a spinon is taken to be 
$Q_\alpha$ on the tetrahedron occupied by it. The \emph{monopole}, also a 
gapped point defect, is the source of the gauge magnetic field, whose presence 
is detected by the Aharonov-Bohm phase of the spinon.
(The monopole is also referred to as \emph{vison} in some 
literature~\cite{Gingras2014}.)

The abundant theoretical predictions~\cite{Hermele2004,Savary2012,Lee2012,Hao2014,ChenGang2017,Kwasigroch2017} on the QSI topological excitations naturally call for numerical scrutiny. Yet, their dynamical properties so far have only been indirectly inferred from the numerical analysis of the ground state or toy models~\cite{Benton2012,WanYuan2016,Henry2014,Stefanos2016,Petrova2015}. Here, we directly address the dynamics problem by unbiased QMC simulation of a QSI model.

\begin{figure*}[htp!]
\includegraphics[width=\linewidth]{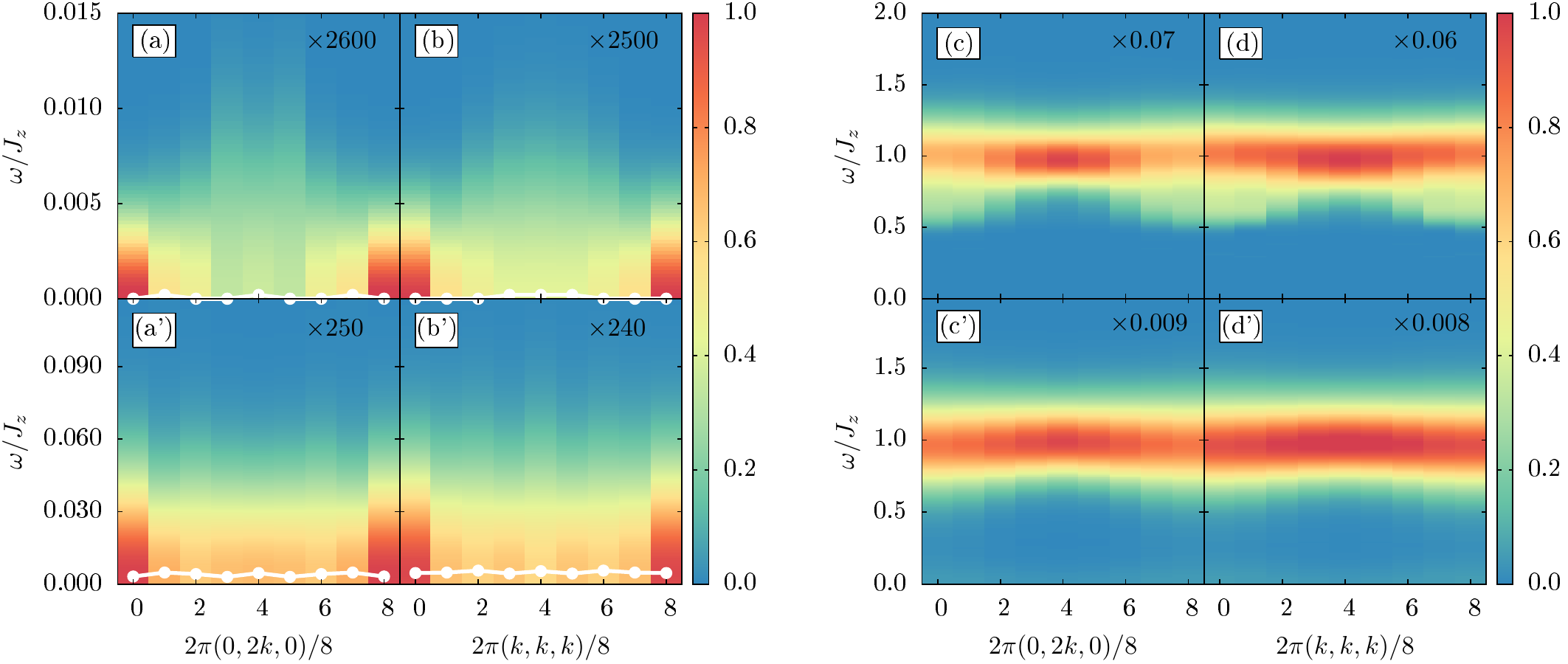}
	\caption{Left panel: Dynamic spin structure factor 
	$S^{zz}(\mathbf{q},\omega)$ obtained from QMC-SAC at temperature $T_2$ 
	(a,b) and $T_3$ (a',b'). White dots mark the position of spectral peaks. 
	Both $T_2$ and $T_3$ are inside the classical spin ice regime. The photon 
	disappears, and the spectra are diffusive. The peak positions in (a',b') 
	are slightly above the horizontal ($\omega=0$) axis. This is likely an 
	artifact due to the uncertainties in the SAC method. Note the $\omega$ axis 
	scale of (a',b') is different from (a,b). Right panel: Dynamic spin 
	structure factor $S^{+-}(\mathbf{q},\omega)$ obtained from QMC-SAC at 
	temperature $T_2$ (c,d) and $T_3$ (c',d'). Comparing to the spectra at 
	$T_1$, the spinon continuum is still present but with narrow and flat 
	dispersion at $T_2$. At $T_3$, the continuum becomes dispersionless.}
\label{fig:fig3}
\end{figure*}

{\it Model} --- We study the \textit{XXZ} model on pyrochlore 
lattice~\cite{Hermele2004},
\begin{align}
\mathcal{H}=\sum_{\langle i,j\rangle}-\jpm(S^+_iS^-_j+h.c.)+\jz S^z_iS^z_j.
\label{eq:model}
\end{align}
Here, $S^{x,y,z}_i$ are the Cartesian components of $S=1/2$ spin operator on site $i$, and the summation is over all nearest-neighbor pairs. $\jz,\jpm>0$ are spin exchange constants.

To set the stage, we briefly review the thermodynamic phase diagram of the 
model in Eq.~\eqref{eq:model}, which has been well established by QMC 
calculations~\citep{Banerjee2008,Shannon2012,Lv2015,Kato2015}. At zero 
temperature, a critical point on the $\jpm/\jz$ axis at 
$J_{\pm,c}/\jz=0.052(2)$~\cite{Lv2015} separates the \textit{XY} ferromagnet 
state ($\jpm>J_{\pm,c}$) and the QSI ground state ($\jpm<J_{\pm,c}$). On the 
QSI side, with fixed $\jpm / \jz$, three regimes exist on the temperature axis. 
At high temperature $k_BT\gg \jz$, the system is in the trivial paramagnetic 
regime with entropy $\mathcal{S}\approx Nk_B\ln 2$, $N$ being the number of 
spins. When $k_BT$ decreases to $O(\jz)$, the system crosses over to a 
classical spin ice (CSI) regime where it \emph{thermally} fluctuates within the 
ice manifold~\cite{Bramwell2001}. Since the number of the spin configurations 
in the ice manifold is exponentially large in $N$, the entropy is still 
extensive: $\mathcal{S}\approx Nk_B\ln(3/2)/2$~\cite{Pauling1935}. As $T$ 
further decreases, the system approaches the QSI regime through a second 
crossover with $\lim_{T\to0}\mathcal{S}=0$. Figure~\ref{fig:fig1}(b) shows the 
entropy $\mathcal{S}$ and specific heat $C$ as a function of $T$ for the 
typical model parameter $\jpm/\jz = 0.046$. The trivial paramagnetic and the 
CSI regimes manifest themselves as plateaux in the entropy, whereas the two 
crossovers appear as two broad peaks in the specific heat respectively located 
at $k_BT/\jz \approx 1$ and $10^{-3}$.

In the ensuing discussion, we set $\jpm/\jz=0.046$ throughout and choose three 
representative temperatures [Fig.~\ref{fig:fig1} (b)]: $k_BT_1=0.001\jz$ (QSI 
regime),  $k_B T_2=0.04\jz$ (CSI regime), and $k_B T_3=0.1\jz$ (close to the 
trivial paramagnetic regime) to perform the QMC simulation and reveal the 
dynamics of topological excitations therein.

{\it Method} --- We numerically solve the model in Eq.~\eqref{eq:model} by using the worm-type, continuous-time QMC algorithm~\cite{Prokofev1998,Prokofev1998b,Lv2015}. As the Hamiltonian $\mathcal{H}$ possesses a global $U(1)$ symmetry, the total magnetization $M^z$ commutes with $\mathcal{H}$. We perform simulation in the grand canonical ensemble where $M^z$ can fluctuate~\cite{Prokofev1998,Note1}. We use a lattice of $8\times 8\times 8$ primitive unit cells with periodic boundary condition.

We characterize the dynamics of topological excitations by dynamic spin structure factors (DSSF),
\begin{subequations}
\begin{align}
S^{+-}_{\alpha\beta}(\mathbf{q},\tau)&=\langle S^{+}_{-\mathbf{q},\alpha}(\tau)S^{-}_{\mathbf{q},\beta}(0) \rangle,\\
S^{zz}_{\alpha\beta}(\mathbf{q},\tau)&=\langle S^{z}_{-\mathbf{q},\alpha}(\tau)S^{z}_{\mathbf{q},\beta}(0)\rangle.
\end{align}\label{eq:corr_def}
\end{subequations}
Here, the imaginary time $\tau$ is related to the real (physical) 
time $t$ by $\tau = it$, and $\alpha,\beta=1,2,3,4$ label the face-center-cubic 
(fcc) sublattices of the pyrochlore lattice. $\langle\cdots\rangle$ stands for 
the QMC ensemble average. 
$S^{\pm}_{\mathbf{q},\alpha}=\sqrt{4/N}\sum_{i\in\alpha}e^{-i\mathbf{q}\cdot\mathbf{r}_i}S^{\pm}_{i}$,
 where the summation is over the fcc sublattice $\alpha$ and $\mathbf{r}_i$ is 
the spatial position of the site $i$. $S^z_{\mathbf{q},\alpha}$ is defined in 
the same vein. 

From the imaginary-time data, we construct the real-frequency spectra 
$S^{+-}_{\alpha\beta}(\mathbf{q},\omega)$ and 
$S^{zz}_{\alpha\beta}(\mathbf{q},\omega)$, which are directly related to 
various experimental probes. They should contain signatures of spinons and 
photons since the spinons are created or annihilated under the action of 
$S^{\pm}_{i}$ operators [Fig.~\ref{fig:fig1}a] and the photons manifest 
themselves in the correlations of $S^z_{i}$ 
operators~\cite{Hermele2004,Benton2012}. The creation or annihilation processes 
of monopoles, however, are not readily related to the local action of the spin 
operators of the \textit{XXZ} model~\cite{Hermele2004}. We therefore expect 
that the signatures of monopoles in DSSF are too weak to allow for direct, 
unambiguous observation. The spectra are constructed by performing the 
state-of-art stochastic analytic continuation 
(SAC)~\cite{Sandvik1998a,Beach2004,Syljuasen2008,Sandvik2015,Qin2017,Shao2017a,Shao2017b}.
 In SAC, we propose candidate real-frequency spectra from the Monte Carlo 
process and fit them to the imaginary time data. Each candidate is accepted or 
rejected according to a Metropolis-type algorithm, where the goodness-of-fit 
$\chi^2$ plays the role of energy.
The final spectrum is the ensemble average of all candidates. A detailed 
account of SAC and its applications in other quantum magnetic systems can be 
found in recent 
Refs.~[\onlinecite{Sandvik2015,Qin2017,Shao2017b,Shu2018,Ma2018,Sun2018}] and 
Sec.~SII 
of the Supplemental material (SM)~\onlinecite{SM2018}. In what follows, we only 
present the trace 
of the DSSF matrix for simplicity: $S^{zz}(\mathbf{q},\omega) =
\sum_{\alpha}S^{zz}_{\alpha\alpha}(\mathbf{q},\omega)$ and $S^{+-}(\mathbf{q},\omega) = \sum_{\alpha}S^{zz}_{\alpha\alpha}(\mathbf{q},\omega)$.

{\it Dynamics in QSI regime} --- We first consider the quantum spin dynamics at $T_1$, which is close to the QSI ground state.

The photon in QSI is analogous to the electromagnetic wave. Since $S^z_i$ is 
akin to the electric field, the QSI photon is visible in the dynamic spin 
structure $S^{zz}(\mathbf{q},\omega)$~\cite{Benton2012}.
Figures~\ref{fig:fig2}(a,b) show QMC-SAC results for 
$S^{zz}(\mathbf{q},\omega)$. The photon appears 
as a single branch of gapless excitation whose excitation energy 
$\omega_\mathbf{q}$ [Figs.~\ref{fig:fig2}(a,b), white dots] vanishes as 
$\mathbf{q}$ approaches the Brillouin Zone (BZ) center. The overall dispersion 
relation qualitatively agrees with the prediction from a simple Gaussian QED 
model (Sec. SIV of SM).  Crucially, the spectral function at $\mathbf{q}=0$ has 
a sharp peak located at zero excitation energy, reflecting the charge 
conservation law present in our system. This is in contrast with a Goldstone 
mode, which would possess a small energy gap in a finite-size system. Although 
the system size is not large enough to unambiguously resolve the linear 
dispersion at small $\mathbf{q}$ from the DSSF, previous QMC works have 
detected the photon linear dispersion from the $T^3$ scaling law of the 
specific heat~\cite{Lv2015,Kato2015}. The photon bandwidth 
$W_{\gamma}\approx5\times10^{-3}\jz$, consistent with the small energy scale of 
the quantum tunneling within the ice manifold $12J^3_\pm/J_z = 
1.17\times10^{-3}J_z$~\cite{Hermele2004}.

The underlying gauge theory structure also manifests itself in the spectral 
weight of the photon. In contrast with a gapless spin wave, whose 
energy-integrated spectral weight would increase as the excitation energy 
$\omega_\mathbf{q}\to 0$, the photon spectral weight 
[Figs.~\ref{fig:fig2}(a,b), pink open circles] decreases as
$\omega_\mathbf{q}\to 0$. This unusual 
behavior is linked to the fact that the electric field ($S^z$) is the canonical 
momentum of the gauge field~\cite{Benton2012}. Furthermore, the ice rule 
dictates that the photon polarization is transverse to the momentum. Here, we 
find that the spectral weight of the transverse component of the DSSF is at 
least 10 times larger than the longitudinal component. The residual 
longitudinal component is attributed to the virtual spinon pairs, which 
temporarily violate the ice rule.

Even though qualitatively agreeing with the predictions from Gaussian QED theory, the QMC-SAC spectra reveal significant photon decay that is not captured by such a simple model. The half width at half maximum at the zone boundary is approximately $3\times10^{-3}\jz$, which is comparable to $W_\gamma$. The large decay rate indicates the strong photon self-energy at temperature $T_1$.

Having numerically observed photon in the dynamic spin structure factor 
$S^{zz}(\mathbf{q},\omega)$, we now turn to spinons. Spinons are visible in 
$S^{+-}(\mathbf{q},\omega)$, which essentially measures the probability for 
producing a pair of spinons with total momentum $\mathbf{q}$ and energy 
$\omega$ [Fig.~\ref{fig:fig1}(a)]. The operator $S^-_{i}$ creates from vacuum 
a charge $Q=1$ spinon in an up tetrahedron and a $Q=-1$ spinon in the 
neighboring down tetrahedron. The action of the XX term in Eq.~\eqref{eq:model} 
hops the spinons to their respective \emph{next} nearest neighbor tetrahedra as 
the term flips \emph{two} spins at each step. Thus, the $Q=1(-1)$ spinon 
propagates in the fcc lattice formed by the center of up (down) tetrahedra.

Figures~\ref{fig:fig2}(c,d) show the dynamic spin structure factors obtained 
by 
QMC-SAC. The spinon pair appear as a broad continuum in the spectra, mirroring 
the fact that the total energy $\omega$ is not a definite function of  
$\mathbf{q}$ as there is no unique way of assigning $\mathbf{q}$ to individual 
spinons. We find a qualitative agreement between the numerically observed 
$S^{+-}(\mathbf{q},\omega)$ and a tight-binding model calculation 
[Figs.~\ref{fig:fig2}(c',d')], where we assume both spinons are free particles 
(see Sec.~SIII of SM for details). Fitting the tight-binding model to the 
QMC-SAC spectra yields a renormalized spinon hopping amplitude $t \approx 
0.031\jz$, which is smaller than the bare value $\jpm = 0.046\jz$ estimated 
from perturbation theory. The bright features in the spectra are attributed to 
the van Hove singularity in the two-spinon density of states~\cite{Van1953}. Our 
results thus suggest the spinon behaves as a coherent quasiparticle with 
renormalized hopping amplitude~\cite{WanYuan2016,Henry2014,Stefanos2016}. 
However, the quantitative difference between the QMC-SAC spectra and the 
tight-binding model underlines the intricate 
interaction between the spinon and the spin background that is beyond the 
simple tight-binding picture~\cite{WanYuan2016,Stefanos2016}.

{\it Dynamics in CSI regime} --- We now study the dynamics of photons and 
spinons at higher temperature $T$. Our results in the QSI regime identify two 
energy scales: the photon bandwidth $W_\gamma \approx 5\times10^{-3}\jz$, and 
the bandwidth of the two-spinon continuum $W_\psi \approx \jz$. We expect the 
photon to disappear at $k_BT>W_\gamma$. Indeed, at $k_BT_2 = 0.04\jz$, we 
observe a diffusive spectra in $S^{zz}(\mathbf{q},\omega)$, whose spectral 
peaks are positioned at zero frequency [Figs.~\ref{fig:fig3}(a,b)]. This 
indicates the fluctuations within the ice manifold has become thermal.

However, as $k_BT_2\ll W_\gamma$, the spinon dynamics remains coherent despite 
the system is in the CSI regime. This is clearly seen in 
$S^{+-}(\mathbf{q},\omega)$, which exhibits a dispersive spinon continuum 
[Figs.~\ref{fig:fig3}(c,d)]. Comparing to the spectra at $T_1$, the continuum 
is narrow in bandwidth and flat in dispersion. Both features suggest spinon 
hopping processes are less coherent at higher temperature. The smaller spectral 
weight of $S^{+-}(\mathbf{q},\omega)$ also indicates overall weaker quantum 
fluctuations.

As the temperature further increases to $k_B T_3 = 0.1\jz$, the thermally 
populated spinons form a dilute gas~\cite{Castelnovo2011}. The spinons now lose 
their quantum character and instead behave as random 
walkers~\cite{Ryzhkin2005,Jaubert2009}. This is reflected in 
$S^{+-}(\mathbf{q},\omega)$ by an almost dispersionless continuum with spectral 
peaks pinned at the classical spinon pair creation energy $\omega=J_z$ 
[Figs.~\ref{fig:fig3}(c',d')]. Meanwhile, $S^{zz}(\mathbf{q},\omega)$ 
[Figs.~\ref{fig:fig3}(a',b')] is even more diffusive comparing to $T_2$. The 
peak width at $T_3$ is about 10 times broader than that at $T_2$, and the 
spectral intensity drops by
a factor of 10 to preserve the sum rule.

{\it Discussion} --- We therefore identify three temperature windows with 
distinct dynamics for the topological excitation. At a very low temperature 
$T$, we numerically observe both coherent gauge photons and fractionalized 
spinons in the DSSF. As $T$ increases above the photon bandwidth, the dynamics 
of the spinon remain coherent, despite that the system is in the CSI regime. As 
$T$ further increases, both spinons and photons cease to exist as quantum 
excitations. 

In the QSI window, while our results show a qualitative agreement with the 
field theory, they suggest significant interaction effects in the dynamics of 
photons and spinons that are not captured by free field theory. In the 
intermediate temperature window, our results point to the interesting 
possibility of observing quantum spinons at a more experimentally accessible 
temperature, which worth further theoretical and numerical exploration.


\begin{acknowledgments}
{\it Acknowledgments} --- We acknowledge Collin Broholm, Bruce Gaulin, Jeff Rau 
and Kate Ross for valuable comments. Y.~J.~D. and Z.~Y.~M. are grateful to Gang 
Chen for enlightening discussions on the zero temperature spinon continuum and 
previous collaborations on related projects. C.~J.~H., Y.~J.~D. and Z.~Y.~M. 
acknowledge the Ministry of Science and Technology of China (under Grants 
2016YFA0300502 and 2016YFA0301604) and the National Natural Science Foundation 
of China under Grants No. 11421092, No. 11574359, No. 11625522, No. 11674370, 
the key research program of the Chinese Academy of Sciences under Grant No. 
XDPB0803, and the National Thousand-Young Talents Program of China. Research at 
Perimeter Institute (YW) is supported by the Government of Canada through the 
Department of Innovation, Science and Economic Development Canada and by the 
Province of Ontario through the Ministry of Research, Innovation and Science. 
We thank the Center for Quantum Simulation Sciences in the Institute of 
Physics, Chinese Academy of Sciences and the Tianhe-1A platform at the National 
Supercomputer Center in Tianjin for their technical support and generous 
allocation of CPU time.
\end{acknowledgments}

%


\setcounter{page}{1}
\setcounter{equation}{0}
\setcounter{figure}{0}
\renewcommand{\theequation}{S\arabic{equation}}
\renewcommand{\thefigure}{S\arabic{figure}}


\begin{widetext}

	\section*{Supplemental Material}

	\centerline{\bf Dynamics of topological excitations in a model quantum spin ice}
	\vskip3mm

	\centerline{C-J. Huang, Y. J. Deng, Y. Wan and Z. Y. Meng}

	\vskip6mm

	In this supplemental material, we provide the technical details concerning the QMC measurements of two types of spin-spin correlation functions (Sec.~SI), the stochastic analytic continuation method (SAC) for extracting real-frequency spectral functions and the analysis of transverse and longitudinal components of the spin spectra (Sec.~SII). The calculations of spectra of spinon and photon in QSI phase with the help of a tight binding model and a lattice QED model respectively can be found in Sec.~SIII and Sec.~SIV.
\end{widetext}

\subsection{SI. Evaluation of dynamical correlation functions}
To obtain spectral information from quantum Monte Carlo simulation, we need to
measure the dynamical correlation function between operators $\hat{O}_1$ and $\hat{O}_2$:
\begin{equation}
	G(\tau)=\langle\hat{O}_1(\tau)\hat{O}_2(0)\rangle.
\end{equation}
In this paper we compute two types of correlation functions,
$S^{+-}_{\alpha,\beta}(\mathbf{q},\tau)$ and $S^{zz}_{\alpha,\beta}(\mathbf{q},\tau)$, as shown in Eq.~\ref{eq:corr_def} of the main text. Within worm-type QMC algorithm, the measurements can be performed both on space and imaginary time axes. In worm algorithm, we have two types of phase spaces, partition function space $\mathcal{Z}$ and Green's function space
$\mathcal{G}$. The measurements of $S^{+-}_{\alpha,\beta}(\mathbf{r},\tau)$ are carried in $\mathcal{G}$ space because it is the set of the $S^{+}S^{-}$ configurations
under path-integral. Every $\mathcal{G}$ configuration contributes one to
corresponding $S^{+-}_{\alpha,\beta}(\mathbf{r},\tau)$ measurements. On the other hand, because $S^z$ is diagonal in the QMC worm algorithm, the $S^{zz}_{\alpha,\beta}(\mathbf{r},\tau)$ measurements are implemented in $\mathcal{Z}$ space. After obtaining the two correlation functions in real space, their momentum dependance can be easily accessed via Fourier transform.

In our algorithm, the imaginary time axis is continuous, 
the measurements of correlation functions are performed on a quadratic grid of imaginary time, instead of the uniform grid, with the set of $\{\tau_i\}$ following $\tau_k=k^2\Delta$ if $\tau_k<\beta/2$, $\tau_k=\beta-(2M-k)^2\Delta$ otherwise, where $k=0,1,2,\cdots,2M$ and $\Delta=\beta/2M^2$. Here we set $M=400$. Due to the symmetry properties, in SAC, only the range $0\le\tau<\beta/2$ has to be considered.

\subsection{\label{SII}SII. SAC and spectra}
\begin{figure*}[t]
\includegraphics[width=0.9\linewidth]{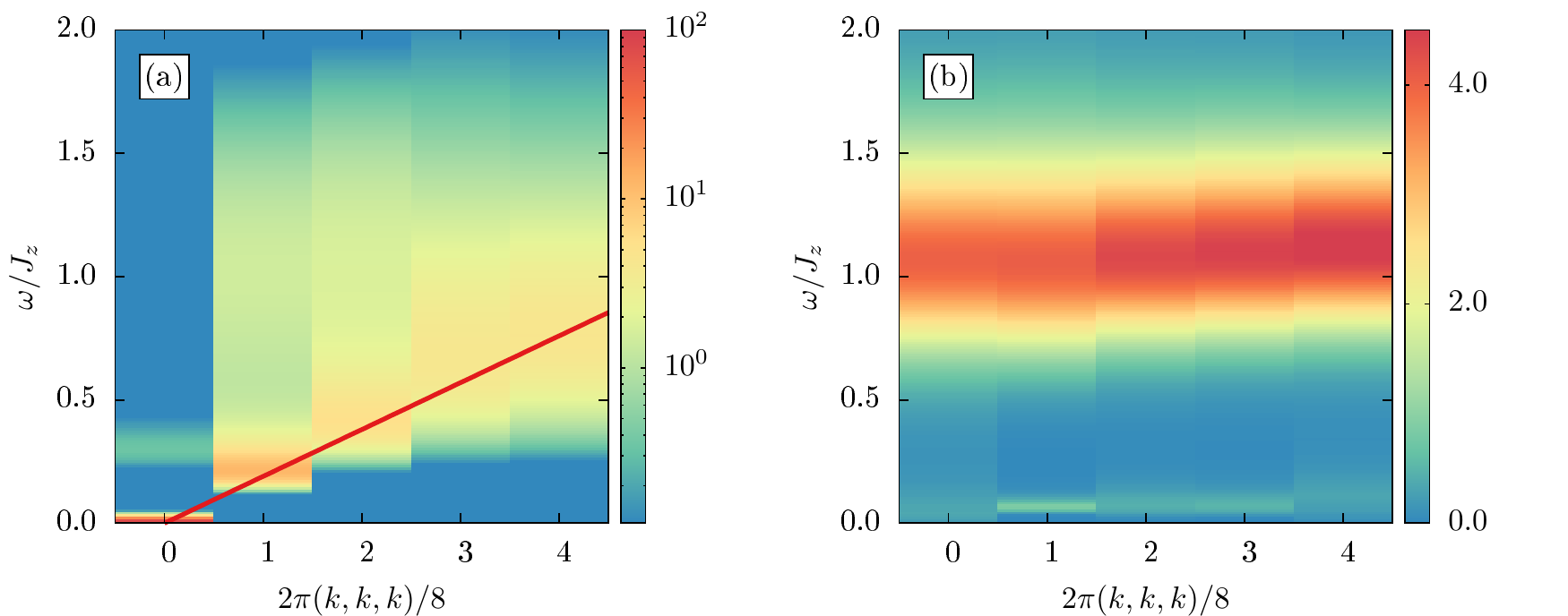}
\caption{\label{l8Jpm0o06lambda} (a) and (b) are the transverse and longitudinal spectra of superfluid phase respectively with the system size $L=8$, $\beta=100$ and $\jpm=0.06$ along the $[1,1,1]$ direction. The solid red line in (a) is a guide to the eye for the spin-wave excitation.}
\end{figure*}

In general, the relationship between imaginary-time correlation function $G(\tau)$ and the
spectral function $A(\omega)$ is as follows:
\begin{equation}
	G(\tau)=\int_{-\infty}^{+\infty} d\omega\ K(\tau,\omega)A(\omega)
\label{eq:inversion}
\end{equation}
where $K(\tau,\omega)$ is the kernel depending on the type of $A(\omega)$. For example, $K(\tau,\omega)=\pi^{-1}e^{-\omega\tau}$ for bosonic spectra. With the help of analytic continuation method, $A(\omega)$ can be obtained from $G(\tau)$. However, analytic continuation is an ill-posed numerical problem~\cite{Jarrell1996,Han2017} as the $G(\tau)$ contains the QMC statistical errors and the kernel contains exponential factors which render the matrix inversion of Eq.~\ref{eq:inversion} very unstable. To overcome these numerical issues, the stochastic analytic continuation (SAC) method has been developed and kept improving over the years~\cite{Sandvik1998a,Beach2004,Syljuasen2008,Sandvik2015,Qin2017,Shao2017a,Shao2017b}.
In this paper, since the spin spectra are all bosonic excitations, we implement the following convention,
\begin{equation}
	\begin{split}
		& \tilde{G}(\tau)=\int_{0}^{+\infty} d\omega\ \tilde{K}(\omega,\tau)\tilde{A}(\omega), \\[8pt]
		\text{with} & \begin{cases}
		\tilde{K}(\omega,\tau) = \dfrac{1}{\pi}\dfrac{e^{-\tau\omega}+e^{-(\beta-\tau)\omega}}{1+e^{-\beta\omega}}\\[10pt]
		\tilde{A}(\omega) = A(\omega)(1+e^{-\beta\omega})
		\end{cases}
	\end{split}
	\label{eq:gka}
\end{equation}
in the SAC employed.

Although in the Refs.~\onlinecite{Sandvik2015,Qin2017,Shao2017a,Shao2017b} there are very detailed description of the SAC, here we would still like to outline the method, for the sake of completeness. The QMC-SAC procedure goes as follows, from QMC measurements, we obtain $G(\tau_i)$ with a set of $\{\tau_{i}\}$ of imaginary time points, and the covariance matrix between the series of $\{G(\tau_i)\}$ describing the Monte Carlo autocorrelation between the data. What should be noted is that for the sake of numerical stability, only these $\{G{(\tau_i)}\}$ data with relative error not larger than $10\%$ are retained for SAC.

Then in SAC, we invert the convolution relation Eq.~\ref{eq:gka} in a Metropolis-type Markov process via Monte Carlo sampling to have $\tilde{A}(\omega)$, and then to obtain the spectral function $A(\omega)$ from $\tilde{A}(\omega)$.

As discussed in the main text, in SAC, a chosen parameterization of the spectrum, for example a large number of $\delta$-functions, is sampled in a Markov Monte Carlo simulation according to the following probability distribution,
\begin{equation}
P(B) \propto \exp(-\chi^2/2\Theta).
\label{eq:SACMC}
\end{equation}
where $B$ means one possible configuration of $A(\omega)$. Here $\chi^2$ is the goodness-of-fit between the $G(\tau)$ and the $\tilde{G}(\tau)$ from proposed $\tilde{A}(\omega)$ in a Monte Carlo instance, and $\Theta$ is a fictitious temperature. Then $\chi^2$ plays the role of an energy and the SAC is converted to a problem in statistical mechanics.
The sampling space is a large number, $N_{\omega}$, of movable $\delta$-functions placed on a frequency grid with a spacing $\Delta_\omega$ sufficiently fine as to be regarded in practice as a continuum (e.g., $\Delta_\omega=10^{-3}\sim 10^{-5}$).
A sweep of the SAC Monte Carlo is consisting of $N_{\omega}$ moves of a single and a couple of $\delta$-functions along the $\omega$ axis. The update is accepted or rejected according to Eq.~\ref{eq:SACMC}. Every sweep will produce a new configuration and the final spectrum is the ensemble average of these configurations. The flow chart of SAC algorithm is summarized in Algorithm.~\ref{tab:sac}.

The choice of $\Theta$ is an important issue in SAC, very low $\Theta$ will freeze the spetrum to metastable $\chi^2$ minimum, while a high $\Theta$ leads to large $\chi^2$, giving rise to poor fits to the QMC data of $G(\tau)$. Here we adopt a simple temperature-adjustment scheme devised in Ref.~\cite{Shao2017a}, where a simulated annealing procedure is first carried out to find the minimum $\chi^{2}_{min}$ and then $\Theta$ is further adjusted so that the average $\chi^{2}$ during the sampling process for collecting the spectrum satisfies the criterion
\begin{equation}
\langle \chi^2 \rangle \approx \chi^{2}_{\text{min}} + \sqrt{2N_\tau},
\end{equation}
where $N_\tau$ is the number of time imaginary time points in the QMC database for $G(\tau)$. This is the standard deviation of the $\chi^2$ distribution, as shown in Algorithm.~\ref{tab:sac}. For more information about the QMC-SAC procedure, the ardious readers are refereed to Refs.~\cite{Qin2017,Shao2017a,Shao2017b}.

Pyrochlore lattice has four sublattices and the $S^{+-}_{\alpha,\beta}(\mathbf{q},\tau)$ and the $S^{zz}_{\alpha,\beta}(\mathbf{q,\tau})$ carry the sublattice index $\alpha,\beta=1,2,3,4$. For the general discussion below, we denote them as $\mathbf{G}(\tau)$, which is a $4\!\times\!4$ matrix
\begin{equation}
	\mathbf{G}(\tau)=
\begin{bmatrix}
	    G^{11} & G^{12} & G^{13} & G^{14}\\
	    G^{21} & G^{22} & G^{23} & G^{24}\\
	    G^{31} & G^{32} & G^{33} & G^{34}\\
	    G^{41} & G^{42} & G^{43} & G^{44}\\
\end{bmatrix}
\end{equation}


\begin{algorithm}[H]
\caption{SAC algorithm}
	\label{tab:sac}
	\algsetup{indent=1em}
	\begin{algorithmic}
		\STATE {\bf BEGIN:} An configuration with $N_\omega$ evenly distributed $\delta$-functions on the frequency axis. $\Theta$ is given a large initial value, e.g.
		$\Theta=10$.
		\STATE {\bf 1. Annealing procedure}
		\LOOP
			\IF {the change of $\chi_{ave}^2$ is smaller than $10^{-3}$}
				\STATE $\chi_{min}^2$=$\chi_{ave}^2$ and exit {\bf loop}
			\ELSE
				\STATE \textit{decrease} $\Theta$ and call {\bf Update Configuration}
			\ENDIF
		\ENDLOOP
		\LOOP
			\IF {$\chi_{ave}^2>\chi_{min}^2 + \sqrt{2N_{\tau}}$}
				\STATE exit {\bf loop}
			\ELSE
				\STATE \textit{increase} $\Theta$ and call {\bf Update Configuration}
			\ENDIF
		\ENDLOOP
		\STATE {\bf 2. Average possible configuration}
		\STATE set an array $A(\omega=0,\omega_1,\cdots,\omega_{max})=0$
		\FOR{$i=1$ to $m$}
			\STATE call {\bf Update Configure}
			\STATE $A(\omega)$=$A(\omega)$+the current configuration
		\ENDFOR\\
		\STATE {\bf Output} $A(\omega)=A(\omega)/m$

		\STATE {\bf *Update Configuration}
		\STATE set $\chi_{ave}^2=0$ and $n$ to an positive integer value
		\FOR{$i=1$ to $n$}
			\STATE move $\delta$-functions along the $\omega$ axis
			\STATE calculate $\chi^2$ of the current configuration\\
			$\chi_{ave}^2$=$\chi_{ave}^2$+$\chi^2$
		\ENDFOR\\
		$\chi_{ave}^2$=$\chi_{ave}^2/n$
	\end{algorithmic}
\end{algorithm}

We can calculate the trace of matrix $\mathbf{G}(\tau)$ and perform SAC on the trace such that we get the whole spectrum directly. But we find, in this way, the resulting spectra are in bad quality with large $\chi^2$. However, since we know there are four branches in the spectra, two are transverse and the other two are longitudinal, we could therefore project $\mathbf{G}(\tau)$ onto transverse and longitudinal channels as
\begin{equation}
	\begin{split}
		G_{\parallel}(\tau) &= \text{Tr}\left[\mathbf{P}(\mathbf{q})\mathbf{G}(\tau)\right]\\
		G_{\perp}(\tau)     &= \text{Tr}\left[\mathbf{Q}(\mathbf{q})\mathbf{G}(\tau)\right]
	\end{split}
\end{equation}
where $\mathbf{P}(\mathbf{q}),\mathbf{Q}(\mathbf{q})$ are the projection matrixes for longitudinal and transverse components, respectively. The forms of $\mathbf{P}(\mathbf{q}),\mathbf{Q}(\mathbf{q})$ are as follows:
\begin{equation}
	\begin{split}
		& \mathbf{P} =\mathbf{V}(\mathbf{V}^t\mathbf{V})^{-1}\mathbf{V}^t \\
		& \mathbf{Q} = \mathbf{1}-\mathbf{P} \\
		\text{with}\quad & \mathbf{V}=\left(
		\begin{matrix}
			{\rm cos}(\kr{1}) & {\rm sin}(\kr{1}) \\
			{\rm cos}(\kr{2}) & {\rm sin}(\kr{2}) \\
			{\rm cos}(\kr{3}) & {\rm sin}(\kr{3}) \\
			{\rm cos}(\kr{4}) & {\rm sin}(\kr{4})
	\end{matrix}\right)
	\end{split}
\end{equation}
where $\mathbf{r}_\alpha$ is the position of an $\alpha$ sublattice site relative to the center of the unit cell.
Then the longitudinal and transverse spectra can be obtained through SAC and 
the complete spectrum is the sum of the two components. The spectra of 
Fig.~\ref{fig:fig2} and Fig.~\ref{fig:fig3} are acquired by this procedure.

\begin{figure*}[t]
\includegraphics[width=\linewidth]{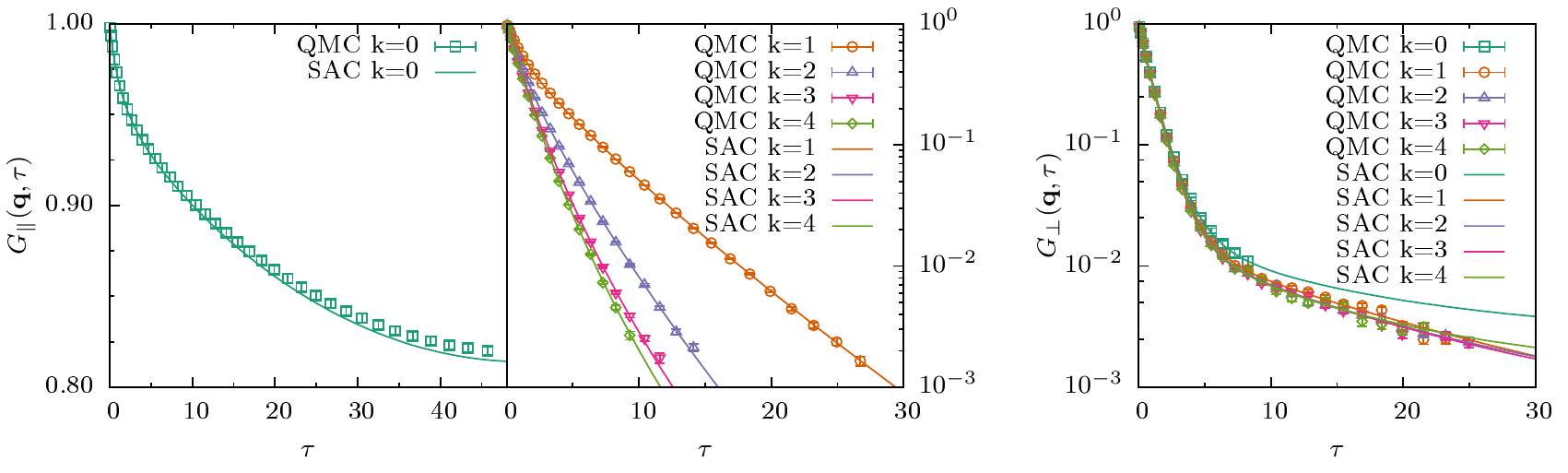}
\caption{\label{l8Jpm0o06lambdag}The comparision of QMC data with the results of SAC in system of size $L=8$, $\beta=100$ and $\jpm=0.06$ along $[1,1,1]$ direction. $\chi^2$ of these results all approximate 1. The meaning of $k$ is the same as in Fig.~\ref{l8Jpm0o06lambda}.}
\end{figure*}

To demonstrate the quality of the spectra after projection, we first measure the spectrum in superfluid phase with $L=8$, $\beta=100$ and $\jpm=0.06$. This is ferromagnetic phase in the spin language. The correlation function $S^{+-}_{\alpha,\beta}(\mathbf{q},\tau)$ is shown along the $[1,1,1]$ momentum direction.
In Fig.~\ref{l8Jpm0o06lambda}, one can see two types of excitations: Goldston 
mode [Fig.~\ref{l8Jpm0o06lambda} (a)]
and the spinon excitation [Fig.~\ref{l8Jpm0o06lambda} (b)]. The former is 
gapless and have a delta
peak in zero momentum points and have dispersion with the increase of momentum
which are consisted with the spin wave theory. The latter means the energy
scale ($\sim J_{z}$) of flipping a single spin which will produce a couple of spinons.

After acquairing the spectrum, we can also transform $\tilde{A}(\omega)$ back to the imaginary time correlation $\tilde{G}(\mathbf{q})$ using Eq.~\eqref{eq:gka}. The results can be directly compared with those measured from QMC, as shown in Fig.~\ref{l8Jpm0o06lambdag}, both for $\tilde{G}_{\parallel}(\mathbf{q},\tau)$ and $\tilde{G}_{\perp}(\mathbf{q},\tau)$, the comparison are in good quality, with $\chi^{2}\sim 1$.

\subsection{SIII. Calculating spinon spectra from a tight binding model}

In this section, we compute the dynamic spin structure factor $S^{+-}_{\alpha\beta}(\mathbf{q},\omega)$ in the QSI regime from a tight-binding model for spinons. In the tight-binding model, we only consider the contribution from two-spinon production process and treat the spinons as free particles.

The dynamic spin structure factor $S^{+-}_{\alpha\beta}(\mathbf{q},t)$ is defined as:
\begin{subequations}
\begin{align}
S^{+-}_{\alpha\beta}(\mathbf{q},t) \equiv \langle0|S^+_{-\mathbf{q},\alpha}(t)S^-_{\mathbf{q},\beta}(0)|0\rangle.
\end{align}
where,
\begin{align}
S^{\pm}_{\mathbf{q},\alpha} \equiv \frac{1}{\sqrt{N_\mathrm{c}}}\sum_{R}S^{\pm}_{R\alpha}e^{-i\mathbf{q}\cdot(\mathbf{R}+\mathbf{r}_\alpha)}.
\end{align}
\end{subequations}
Here $N_\mathrm{c}$ is the number of primitive unit cells in a pyrochlore lattice. $R$ labels the primitive unit cell. $\alpha,\beta$ label the four sublattices, running from 1 to 4. $\mathbf{R}$ is the position of the center of the unit cell. $\mathbf{r}_\alpha$ is the position of an $\alpha$ sublattice site relative to the center of the unit cell.

Acting $S^-_{R\alpha}$ on a spin ice state creates a $Q=1$ spinon and an $Q=-1$ spinon. The $Q=1$ spinon is located at $\mathbf{R}$, which is the center of an up tetrahedron, whereas the $Q=-1$ spinon is located at $\mathbf{R}+2\mathbf{r}_\alpha$, which is the center of a neighboring down tetrahedron. We therefore may approximate $S^-_{R\alpha}\approx a^\dagger_\mathbf{R} b^\dagger_\mathbf{R+2r_\alpha}$, where $a^\dagger_\mathbf{R} (b^\dagger_\mathbf{R'})$ creates a $Q=1$ ($Q=-1$) spinon in an up (down) tetrahedron at $\mathbf{R} (\mathbf{R'})$. Crucially, the $Q=1$ ($Q=-1$) spinon propagates in the face-centered-cubic (FCC) lattice formed by the up (down) tetrahedra. Equipped with this approximation, we find
\begin{align}
S^-_\mathbf{q\alpha} \approx \frac{1}{\sqrt{N_\mathrm{c}}}\sum_{\mathbf{k}} a^\dagger_\mathbf{k} b^\dagger_{-\mathbf{q}-\mathbf{k}}  e^{i(\mathbf{q}+2\mathbf{k})\cdot\mathbf{r}_\alpha} 
\end{align}

Plugging the above into the definition of $S^{+-}_{\alpha\beta}(\mathbf{q},t)$ and using the Wick theorem, we find,
\begin{align}
S^{+-}_{\alpha\beta}(\mathbf{q},t) &\approx \frac{1}{N_\mathrm{c}}\sum_{\mathbf{k}} e^{-i\epsilon_\mathbf{k}t}e^{-i\epsilon_{-\mathbf{q}-\mathbf{k}}t}\nonumber\\
&\times e^{-i(\mathbf{q}+2\mathbf{k})\cdot(\mathbf{r}_\alpha-\mathbf{r}_\beta)}
\end{align}
Here, we have approximate the spinon propagator by a free particle propagator. $\epsilon_\mathbf{k}$ is the dispersion relation for the spinons. The dispersion relations for $Q=\pm1$ spinons are identical due to the time-reversal symmetry and the pyrochlore site-inversion symmetry. Switching to the frequency domain,
\begin{align}
S^{+-}_{\alpha\beta}(\mathbf{q},\omega) &\approx \frac{1}{N_\mathrm{c}}\sum_\mathbf{k} \delta(\omega-\epsilon_\mathbf{k}-\epsilon_{-\mathbf{q}-\mathbf{k}})\nonumber\\
&\times e^{-i(\mathbf{q}+2\mathbf{k})\cdot(\mathbf{r}_\alpha-\mathbf{r}_\beta)}.
\end{align}
In particular, the diagonal components
\begin{align}
S^{+-}_{\alpha\alpha}(\mathbf{k},\omega) \approx \frac{1}{N_\mathrm{c}}\sum_\mathbf{k} \delta(\omega-\epsilon_\mathbf{k}-\epsilon_{\mathbf{k}-\mathbf{k}}),
\end{align}
is simply proportional to the two-spinon density of states within this approximation.

To model the dispersion relation of the spinons, we consider the following tight-binding dispersion in FCC lattice:
\begin{align}
\epsilon_\mathbf{k} &= \frac{J_z}{2}-4t(\cos\frac{k_x}{2}\cos\frac{k_y}{2}+\cos\frac{k_x}{2}\cos\frac{k_z}{2} \nonumber\\
& +\cos\frac{k_y}{2}\cos\frac{k_z}{2}).
\end{align}
where $t>0$ is the effective hopping amplitude. The constant $J_z/2$ is the on-site energy cost for creating a single spinon. To determine $t$, we fit the two-spinon band width $32t$ to the observed width of the two-spinon continuum in QMC-SAC, which yields $t \approx 0.031J_{z}$.

Finally, we phenomenologically incorporate the finite life time effect by broadening the Dirac delta function to Lorentzian:
\begin{align}
\delta(x)\to \frac{1}{\pi}\frac{\gamma}{x^2+\gamma^2},
\end{align}
where $\gamma$ may be interpreted as the spinon-pair decay rate. In practice, we set $\gamma = 2t$.

\subsection{SIV. Calculating photon spectra from a Gaussian QED model}

In this section, we compute the dynamic spin structure factor $S^{zz}_{\alpha\beta}(\mathbf{q},\omega)$ from a Gaussian QED model. Our treatment essentially follows that of \cite{Benton2012}.

We consider the following lattice QED Hamiltonian in the Coulomb gauge:
\begin{align}
H_\mathrm{QED}/\Lambda = \frac{1}{2}\sum_{r}E^2_r+\frac{u^2}{2}\sum_{p}(\mathrm{curl}_p A)^2
\end{align}
Here $u$ is a phenomenological, dimensionless parameter. $\Lambda$ sets the overall energy scale. The first summation is over all pyrochlore sites. The second summation is over all hexagonal rings $p$ of the pyrochlore lattice. $\mathrm{curl}_p$ is the lattice curl associated with $p$. $E_r$ is the electric flux, whereas $A_{r}$ is the gauge potential. They obey the canonical commutation relation: $[A_r,E_{r'}] = i\delta_{r,r'}$. The Hilbert space of the QED model is subject to the Gauss law constraint and the Coulomb gauge condition:
\begin{equation}\begin{split}
\sum_{r\in\alpha}E_r = 0,\quad\forall\alpha,\\
\sum_{r\in\alpha}A_r = 0,\quad\forall\alpha,
\end{split}\end{equation}
where the summation is over a tetrahedron $\alpha$. The first identity is the Gauss law in vacuum, i.e. the divergence of electric field is zero everywhere. The second identity follows from our choice of the Column gauge.

We perform a lattice Fourier transform to diagonalize the above Hamiltonian: $A_{r\in\alpha} = \sqrt{4/N}\sum_{\mathbf{q}}A_{\mathbf{q},\alpha}e^{i\mathbf{q}\cdot\mathbf{r}}$, and $E_{r\in\alpha} = \sqrt{4/N}\sum_{\mathbf{q}}E_{\mathbf{q},\alpha}e^{-i\mathbf{q}\cdot\mathbf{r}}$. The Hamiltonian then may be cast in matrix form:
\begin{align}
H_\mathrm{QED}/\Lambda = \sum_\mathbf{q}\frac{1}{2}E^\dagger_\mathbf{q} E^{\phantom\dagger}_\mathbf{q}+\frac{u^2}{2}A^\dagger_\mathbf{q}Z^\dagger_{\mathbf{q}}Z^{\phantom\dagger}_{\mathbf{q}}A^{\phantom\dagger}_\mathbf{q}.
\end{align}
Here, $A_{\mathbf{q}} = (A_{\mathbf{q},1},A_{\mathbf{q},2},A_{\mathbf{q},3},A_{\mathbf{q},4})^t$ is the $4\times1$ column vector. $E_{\mathbf{q}}$ is defined in the same vein. $4\times4$ matrix $Z_\mathbf{q}$ is given by:
\begin{align}
Z_\mathbf{q}= \left(\begin{array}{cccc}
0 & \zeta_{34} & \zeta_{42} & \zeta_{23}\\
\zeta_{43} & 0 & \zeta_{14} & \zeta_{31}\\
\zeta_{24} & \zeta_{41} & 0 & \zeta_{12}\\
\zeta_{32} & \zeta_{13} & \zeta_{21} & 0
\end{array}\right),
\end{align}
which is the Fourier transform of the lattice curl. $\zeta_{\alpha\beta} = 2i\sin(\mathbf{q}\cdot(\mathbf{r}_\alpha-\mathbf{r}_\beta))$. The constraints are:
\begin{align}
V^{\dagger}_{\mathbf{q},\mu = 1,2}A_\mathbf{q} = V^{\dagger}_{\mathbf{q},\mu = 1,2}E_\mathbf{q} = 0,
\end{align}
where $V_{\mathbf{q},1} = (\cos(\mathbf{q}\cdot\mathbf{r}_1),\cos(\mathbf{q}\cdot\mathbf{r}_2),\cos(\mathbf{q}\cdot\mathbf{r}_3),\cos(\mathbf{q}\cdot\mathbf{r}_4))^t$, and $V_{\mathbf{q},2} = (\sin(\mathbf{q}\cdot\mathbf{r}_1),\sin(\mathbf{q}\cdot\mathbf{r}_2),\sin(\mathbf{q}\cdot\mathbf{r}_3),\sin(\mathbf{q}\cdot\mathbf{r}_4))^t$. Crucially, $V_{\mathbf{q},n=1,2}$ are also in the kernel of $Z_\mathbf{q}$: $Z_\mathbf{q}V_{\mathbf{q},n=1,2} = 0$.

The Fourier-transformed $Z_{\mathbf{q}}$ now may be readily diagonalized. We define the photon creation / annihilation operators $a^{\dagger}_{\mathbf{q}\lambda},a^{\phantom{\dagger}}_{\mathbf{q}\lambda}$ by (in matrix form):
\begin{align}
A_{\mathbf{q}} &= \frac{1}{\sqrt{2\omega_\mathbf{q}}}\sum_{\lambda=1,2} (U_{\mathbf{q},\lambda}a_{\mathbf{q},\lambda}+U^\ast_{-\mathbf{q},\lambda}a^\dagger_{\mathbf{q},\lambda})\nonumber\\
E_{\mathbf{q}} &= i\sqrt{\frac{\omega_\mathbf{q}}{2}}\sum_{\lambda=1,2}(U^\ast_{\mathbf{q},\lambda}a^\dagger_\mathbf{q,\lambda}-U_{-\mathbf{q},\lambda}a_{-\mathbf{q}}).
\end{align}
Here $\lambda= 1,2$ labels the two eigenvectors $U_{\mathbf{q},\lambda}$ of $Z_{\mathbf{q}}$ with non-zero eigenvalues $\pm\omega_\mathbf{q}$. The explicit expression of photon dispersion relation $\omega_\mathbf{q}$ is given by,
\begin{align}
\omega_\mathbf{q} = 2u\Lambda\sqrt{\sum_{\alpha<\beta}\sin^2(\mathbf{q}\cdot\mathbf{d}_{\alpha\beta})}.
\end{align}

The dynamic spin structure factor $S^{zz}_{\alpha\beta}(\mathbf{q},t)$ is related to the correlation function of the electric field:
\begin{align}
S^{zz}_{\alpha\beta}(\mathbf{q},t) &\sim C_{\alpha\beta}(\mathbf{q},t) = \langle 0| E_{-\mathbf{q}\alpha}(t)E_{\mathbf{q}\beta}(0)|0\rangle \nonumber\\
& = \frac{\omega_\mathbf{q}}{2}[(n_\mathbf{q}+1)e^{-i\omega_\mathbf{q}t}+n_\mathbf{q}e^{i\omega_\mathbf{q}t}]\sum_{\lambda=1,2}U_{\mathbf{q},\alpha\lambda}U^\ast_{\mathbf{q},\beta\lambda} \nonumber\\
& = \frac{\omega_\mathbf{q}}{2}[(n_\mathbf{q}+1)e^{-i\omega_\mathbf{q}t}+n_\mathbf{q}e^{i\omega_\mathbf{q}t}]P_{\alpha\beta}.
\end{align}
Here $P_{\alpha\beta}$ is the projection matrix that projects onto the cokernel of $Z_\mathbf{q}$, i.e. the directions transverse to the crystal momentum $\mathbf{q}$. $n_\mathbf{q}$ is the Bose factor. Switching to frequency domain, we find:
\begin{align}
& S^{zz}(\mathbf{q},\omega) = \sum_{\alpha} S^{zz}_{\alpha\alpha}(\mathbf{q},t) \nonumber\\
& \sim \omega_\mathbf{q}[(n_\mathbf{q}+1)\delta(\omega-\omega_\mathbf{q})+n_\mathbf{q}\delta(\omega+\omega_\mathbf{q})].
\end{align}

The QED model depends on two unknown parameters $u$ and $\Lambda$. Their product $u\Lambda$ controls the bandwidth of the photon. We adjust the value of $u\Lambda$ such that the photon bandwidth matches the value measured from QMC-SAC.
\end{document}